\documentclass[twocolumn,           
               showpacs,            
               nopreprintnumbers,     
               aps,                 
               prd,          	    
               letterpaper,             
               groupeaddress,      
               nofootinbib,         
               tightenlines,        
               floats,floatfix      
               ]{revtex4-1}

\usepackage{graphicx}
\usepackage{dcolumn,soul}
\usepackage{bm}
\usepackage{amsmath}
\usepackage{amsfonts,amssymb}
\usepackage{color}
\usepackage{enumerate}
\usepackage{subfigure}

\begin{document}

\title{Noncommutative Quantum Cosmology for the FLRW model in Unimodular Gravity}

\author{J. A. Astorga-Moreno$^{1}$}
\email{jesus.astorga@cinvestav.mx}

\author{Miguel A. Garc\'{\i}a-Aspeitia$^{2}$}
\email{angel.garcia@ibero.mx}

\author{E. A. Mena-Barboza$^{3}$}
\email{eri.mena@academicos.udg.mx}

\affiliation{$^1$Departamento de F\'isica, Centro de Investigaci\'on y Estudios Avanzados del IPN,\\ Apartado Postal 14-740, 07000, CDMX, M\'exico.}

\affiliation{$^2$Depto. de Física y Matemáticas, Universidad Iberoamericana Ciudad de México, Prolongación Paseo \\ de la Reforma 880, México D. F. 01219, México}

\affiliation{$^3$Centro Universitario de la Ci\'enega, Universidad de Guadalajara\\
Ave. Universidad 1115, Ed. de Investigaci\'on y Tutor\'ias, C.P. 47820 Ocotl\'an, Jalisco, M\'exico.}

\date{\today}
\begin{abstract}

It is well known that classical Unimodular Gravity (UG) is equivalent to General Relativity, but is not clear the equivalence at the quantum level. Here we exhibit, in the UG framework, the Quantum Cosmology (QC) for a Friedmann-Lemaitre-Robertson-Walker (FLRW) universe, considering an inflationary scenario and also introducing a deformation on the commutation relations for the minisuperspace variables (NCC). 

 \end{abstract}
 \maketitle


\section{Introduction} \label{intro}

The Universe acceleration is one of the most profound mysteries of the Universe evolution, being observed first by Supernovaes \cite{1} and confirmed by the acoustic peaks of Cosmic Microwave Background Radiation (CMB) \cite{2}. The straightforward explanation of the Universe acceleration can be through the presence of an effective cosmological constant (CC), whose role fit extraordinarily well with diverse cosmological observations \cite{3}. However, despite its benefits, the CC is afflicted with several problems. Among the most vicious, is the problem when we try to quantify the quantum vacuum oscillations, to explain the number obtained by observations \cite{4,5}. The theoretical prediction is 120 orders of magnitude of difference with respect to the value given by the observations. Another problem that afflicts the CC is the \textit{coincidence problem} due that the domination epoch of CC is around $z\sim0.7$, i.e. in late time of the Universe evolution. These are some reason of why the community dedicates time to explore other paths to tackle the problem. The most studied are for example the dynamical dark energy (DDE), like the Chaplygin gases \cite{6}, the emergent DE that comes from the presence of scalar fields \cite{7}. Other alternatives to affront the problem of Universe acceleration are, for example, brane theories \cite{8,9,10}, $f(R)$ models \cite{11}, unimodular gravity (UG) \cite{12,13}, which implies modifications to General Relativity (GR) in order to obtain a natural late acceleration without the addition of an exotic fluid (see \cite{14} for a compilation of models).

On the other hand, one of the best candidates to tackle the cosmic acceleration, is the previously mentioned UG \cite{15}, whose main characteristic is that scalar density is considered a constant in contrast with GR. The imposition of this, generates a gravitational field equation with the characteristic that it is trace-free and the energy momentum tensor is not conserved at least if this is imposed as an extra hypothesis. If we add the energy-momentum conservation, the traditional Einstein field equations are recovered with the apparition of a CC with non gravitating characteristics due that it is an integration constant, having the capability to add the expected value, obtained from observations \cite{12,13}.

Recently, at the quantum level, the UG is also studied in order to generate a coherent quantum theory of gravity. We know that classically the UG theory is equivalent to GR when it is imposed the condition $\nabla^{\nu}T_{\mu\nu}=0$, however differs when it comes to quantum theory \cite{16}. Is in this vein that we study the quantum cosmology scenario through the canonical quantization. We start this performance using the FLRW metric with a scalar field with an exponential potential, it is derived and exhibit relations in a UG framework through the Wheeler-DeWitt (WDW) and  the Einstein-Hamilton-Jacobi (EHJ) equations. The results are  presented via a semiclassical limit, in order to see the evolution through the temporal variable. This task is reached through the introduction of noncommutativity between minisuperspace coordinates.

The paper outline is as follows: In Sec. \ref{Sec:wdw}, we present the mathematical formalism of unimodular gravity and the WDW equation. In Sec. \ref{Sec:UGSEM} it is deduced the semiclassical approximation for the WDW equation in UG context. Finally, Sec. \ref{CO} is dedicated to conclusions and discussions. From here, we will use natural units in which $c=\hbar=k_{B}=1$.

\section{WDW equation in Unimodular gravity}\label{Sec:wdw}

Unimodular gravity can be followed from the action
\begin{equation}
    S=\int dx^4\sqrt{-g}\left(\frac{1}{2\kappa}R-\mathcal{L}_{matt}\right),
\end{equation}
where $\kappa\equiv8\pi G$, $\mathcal{L}_{matt}$, corresponds to the matter Lagrangian and $R$ is the usual Ricci scalar\footnote{For a discussion of the equivalence between metrics in unimodular coordinates and other systems for a FRLW Universe, see \cite{20,21}.}. In Unimodular gravity context, the scalar density is usually constrained in the form $\sqrt{-g}=\epsilon_0$, hence, the action can be manipulated in order to found the following field equations
\begin{equation}
    R_{\mu\nu}-\frac{1}{4}g_{\mu\nu}R=8\pi G\left(T_{\mu\nu}-\frac{1}{4}g_{\mu\nu}T\right),
\end{equation}
where the energy momentum tensor is defined as usual, this happens because the combination of $T_{\mu\nu}-g_{\mu\nu}T/4$ is independent of what energy-momentum tensor we are using\cite{17}.\newline
Now, the line element for a homogeneous an isotropic universe is
\begin{equation}
ds^2=-N^2(t)dt^2+a^2(t)\Big[\frac{dr^2}{1-kr^2}+r^2(d\theta^2+\sin^2\theta d\varphi^2)\Big],
\end{equation}
where $a$ is the scale factor, $N$ the lapse function and $k$ the curvature constant that takes the values $0,1,-1$ that correspond to a flat, open and closed universe, respectively. In references \cite{18,19} appears a relevant analysis for the flat case in the early Universe, considering 
\begin{equation}
S=S_{g}+S_{\phi}=\int dx^4\sqrt{-g}\Big[R-2\Lambda-\frac{(\dot{\phi})^2}{2N^2}-V(\phi)\Big],\label{accion}
\end{equation}
taking in mind that $\phi$ is a scalar field endowed with a scalar potential $V$, $\Lambda$ is the cosmological constant. Now, the metric determinant is restricted to satisfy the unimodular condition $\sqrt{-g}=\epsilon_0$, where $\epsilon_0$ is a fixed scalar density, then the unimodular action $S_{UN}$ can be written as 
\begin{eqnarray}
S_{UN}=\int^{t_1}_{t_0}dt\frac{\epsilon_0}{N}\Big[-\frac{12(\dot{a})^2}{a^2N}-\frac{6\dot{a}\dot{N}}{aN^2}-\frac{(\dot{\phi})^2}{2N}-\nonumber\\
N(V+2\Lambda+6ka^{-2})\Big],
\end{eqnarray}
making the ansatz $\dot{a}/a=\dot{\alpha}$ the Lagrangian becomes
\begin{eqnarray}
\mathcal{L}=\frac{\epsilon_0}{N^2}\Bigg[-\dot{\alpha}\Big(12\dot{\alpha}+\frac{6\dot{N}}{N}\Big)-\frac{(\dot{\phi})^2}{2}-
N^2\Big(V(\phi)+\nonumber\\
2\Lambda+
6ke^{-2\alpha}\Big) \Bigg],\label{lag}
\end{eqnarray}
obtaining the canonical momenta and fixing $N=e^{-3\alpha}$, we are in a position to write the corresponding Hamiltonian
\begin{equation}
\mathcal{H}=\frac{e^{-6\alpha}}{\epsilon_0}\Bigg[\frac{\Pi^2_{\alpha}}{24}-\frac{\Pi^2_{\phi}}{2}+\epsilon_0e^{6\alpha}\Big(V(\phi)+2\Lambda+6ke^{-2\alpha} \Big)\Bigg].
\end{equation}
Replacing $\Pi_{q^\mu}=-i\partial_{q^\mu}$, here $q^\mu=(\alpha,\phi)$, 
and $\mathcal{H}=0$ implies the WDW equation in the unimodular framework (UGWDW)
\begin{equation}
\Box\psi+\epsilon_0e^{6\alpha}\Big(V(\phi)+2\Lambda+6ke^{-2\alpha} \Big)\psi=0,\label{wdwp}
\end{equation}
where
\begin{equation}
\Box\equiv-\frac{1}{24}\partial^2_{\alpha}+\frac{1}{2}\partial^2_{\phi},
\end{equation}
is the two dimensional modified D'Alambertian. Applying the following transformation between coordinates
\begin{equation}
u=-6\alpha+\lambda\phi, \quad v=2\alpha, \label{trans}
\end{equation}
if $\lambda$ is a parameter, then using $V=V_0e^{-\lambda\phi}$, we get\footnote{This potential represents a chaotic inflation model.} in \eqref{wdwp} 
\begin{eqnarray}
\partial^2_{u}\psi-\frac{1}{3(\lambda^2-3)}\partial^2_{v}\psi+\nonumber\\
\frac{2\epsilon_0V_0e^{-u}}{\lambda^2-3}\psi+
\frac{2\epsilon_0}{\lambda^2-3}\Big(2\Lambda e^{3v}+6ke^{2v} \Big)\psi=0.\label{wdwp1}
\end{eqnarray}
Considering $\psi=X(u)Y(v)$, we have the following two equations
\begin{subequations}
\begin{eqnarray}
&&\frac{d^2 X}{du^2}+(\beta e^{-u}-\eta^2)X=0,\label{sub1} \\
&&\frac{d^2 Y}{dv^2}-\Big(12\epsilon_0\Lambda e^{3v} +36k\epsilon_0e^{2v}+3(\lambda^2-3)\eta^2\Big)Y\nonumber\\&&=0,
\label{sub2}
\end{eqnarray}
\end{subequations}
with $\eta^2$ the separation constant and $\beta=\frac{2\epsilon_0V_0}{\lambda^2-3}$. The equation \eqref{sub1} can be treated as a Bessel one, resulting
\begin{equation}
X=
\left\lbrace
\begin{array}{ll}
c_0J_{2\eta}\Big(2\sqrt{\beta}e^{-\frac{u}{2}}\Big)+
c_1Y_{2\eta}\Big(2\sqrt{\beta}e^{-\frac{u}{2}}\Big),\quad  \lambda>\sqrt{3}\\\\
c_0J_{2\eta}\Big(2i\sqrt{\beta}e^{-\frac{u}{2}}\Big)+
c_1Y_{2\eta}\Big(2i\sqrt{\beta}e^{-\frac{u}{2}}\Big), \quad  \lambda<\sqrt{3}. 
\end{array}
\right.
\end{equation}
In the flat case \eqref{sub2} can be transformed as a modified Bessel equation then, taking $\kappa=\frac{2\sqrt{(\lambda^2-3)}\eta}{\sqrt{3}}$, we have
\begin{equation}
Y=
\left\lbrace
\begin{array}{ll}
c_0I_{\kappa}\Big(\frac{4\sqrt{\epsilon_0\Lambda}}{\sqrt{3}}e^{\frac{3v}{2}}\Big)+
c_1K_{\kappa}\Big(\frac{4\sqrt{\epsilon_0\Lambda}}{\sqrt{3}}e^{\frac{3v}{2}}\Big),\quad \lambda>\sqrt{3}\\\\
c_0I_{i\kappa}\Big(\frac{4\sqrt{\epsilon_0\Lambda}}{\sqrt{3}}e^{\frac{3v}{2}}\Big)+
c_1K_{i\kappa}\Big(\frac{4\sqrt{\epsilon_0\Lambda}}{\sqrt{3}}e^{\frac{3v}{2}}\Big), \quad  \lambda<\sqrt{3}. 
\end{array}
\right.\label{yconm}
\end{equation}
The product $XY=\psi_{\eta}$ forms a family of solutions because have a dependence on the parameter $\eta$, thus the general solution, using a weighting function $F(\eta)$, can be put as 
\begin{equation}
\psi=\int{F(\eta)\psi_{\eta}d\eta}.
\end{equation}
Now we propose a deformation in the variables that obey kind of commutation relation as in noncommutative quantum mechanics\cite{gamboa}, in our work, this noncommutativity can be applied to a quantum cosmological model, where $q_{nc}^\mu=q^\mu\pm\frac{\theta \Pi_{q^\mu}}{2}$ the noncommutative in coordinates unimodular gravity Wheeler-DeWitt equation (NCC-UGWDW), the theta parameter is an uncertainty between the variables, analogous to the Heisenberg uncertainty principle but on the spatial or mini-superspace variables
\begin{equation}
\Box\psi_{nc}+\epsilon_0e^{6\alpha_{nc}}\Big(V(\phi_{nc})+2\Lambda+6ke^{-2\alpha_{nc}} \Big)\psi_{nc}=0,\label{nccwdwp0}
\end{equation}
where we have proposed a particular ansatz in the configuration coordiantes
\begin{equation}
 \alpha_{nc}=\alpha+\frac{\Pi_{\phi}\theta}{2}, \qquad  \phi_{nc}=\phi-\frac{\Pi_{\alpha}\theta}{2},\label{bs}
\end{equation}
which satisfy the deformed algebra
\begin{equation}
 \{\phi_{nc}, \alpha_{nc} \}=\theta, \qquad \{\alpha_{nc}, \Pi_{\alpha}\}=\{\phi_{nc},\Pi_{\phi} \}=1, \nonumber 
\end{equation}
\begin{equation}
  \{\Pi_{\phi},\Pi_{\alpha} \}=0,
\end{equation}
we choose the momenta in both spaces are the same $\Pi^{nc}_{\alpha}=\Pi_{\alpha}$ and via the above transformations we have in \eqref{nccwdwp0}
\begin{eqnarray}
\Box\psi_{nc}+\epsilon_0(V_0e^{6\alpha-\lambda\phi}e^{-i\theta(3\partial_{\phi}+\lambda\partial_{\alpha})}+\nonumber\\
2\Lambda e^{6\alpha}e^{-3i\theta\partial_{\phi}}
+6ke^{4\alpha}e^{-2i\theta\partial_{\phi}})\psi_{nc}=0.
\label{nccwdwp}
\end{eqnarray}
and using the properties
\begin{align}
e^{\eta(A+B)}&=e^{-\eta^2[A,B]}e^{\eta A}e^{\eta B},\nonumber\\
e^{i\theta\partial_x}e^{\eta x}&=e^{i\eta\theta}e^{\eta x},\label{relat}
\end{align}
together with \eqref{trans} and the anzats $\psi_{nc}=e^{\pm ip_1u}Y_{nc}(v)$, gives 
\begin{eqnarray}
\frac{d^2 Y_{nc}}{dv^2}-\Big(12\epsilon_0\Lambda e^{3v} e^{\pm 3p_1\lambda\theta}+36k\epsilon_0e^{2v}e^{\pm 2p_1\lambda\theta} +\nonumber\\
3(\lambda^2-3)p_1^2\Big)Y_{nc}=0.\label{noc}
\end{eqnarray}
then when $k=0$ and making $w=\frac{   4\sqrt{\epsilon_0\Lambda}e^{\frac{3v}{2}} e^{   \frac{\pm3ip_1\lambda\theta}{2}   }}{\sqrt{3}}$ results a modified Bessel equation
with $\kappa=\frac{2\sqrt{\lambda^2-3}p_1}{\sqrt{3}}$, getting for $Y_{nc}$ when $\lambda>\sqrt{3}$ and $\lambda<\sqrt{3}$, respectively 
\begin{equation}
\left\lbrace
\begin{array}{ll}
c_0I_{\kappa}\Big(\frac{   4\sqrt{\epsilon_0\Lambda}e^{\frac{-3\bar{v}}{2}} e^{   \frac{\pm3p_1\lambda\theta}{2}   }}{\sqrt{3}}\Big)+
c_1K_{\kappa}\Big(\frac{   4\sqrt{\epsilon_0\Lambda}e^{\frac{-3\bar{v}}{2}} e^{   \frac{\pm3p_1\lambda\theta}{2}   }}{\sqrt{3}}\Big),\\\\
c_0I_{i\kappa}\Big(\frac{   4\sqrt{\epsilon_0\Lambda}e^{\frac{-3\bar{v}}{2}} e^{   \frac{\pm3p_1\lambda\theta}{2}   }}{\sqrt{3}}\Big)+
c_1K_{i\kappa}\Big(\frac{   4\sqrt{\epsilon_0\Lambda}e^{\frac{-3\bar{v}}{2}} e^{   \frac{\pm3p_1\lambda\theta}{2}   }}{\sqrt{3}}\Big).  
\end{array}
\right.
\end{equation}
under the diffeomorphism $v \leftrightarrow -\bar{v}$ and also recovering \eqref{yconm} when $\theta \to 0$. Proceeding similarly, the commutative and NCC cases when $\Lambda=0$, $k\neq 0$ and $\Lambda=0$, $k=0$ also can be studied. 

\begin{figure}[ph]
\centerline{
\includegraphics[width=0.45\textwidth]{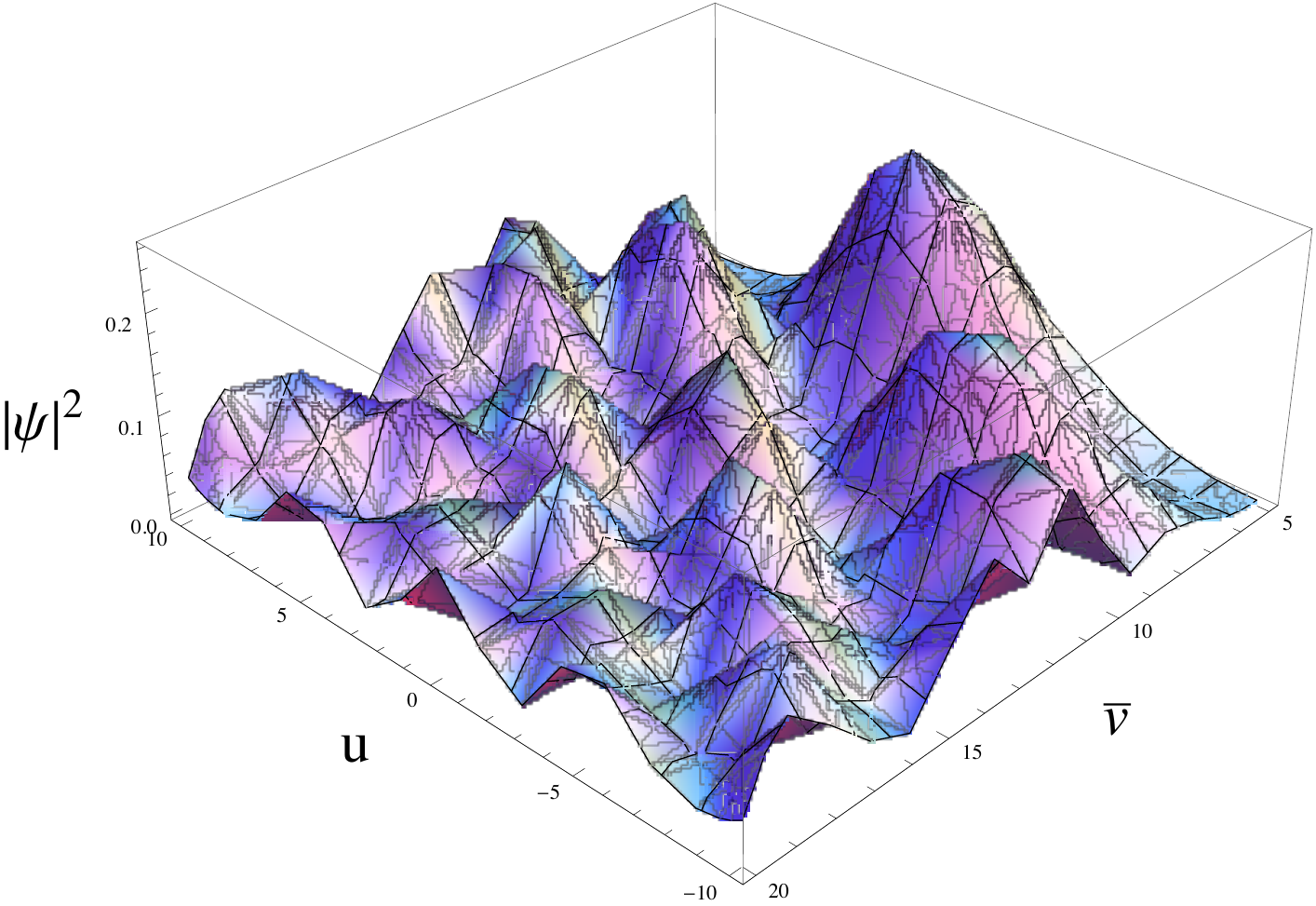}\label{d4}}
\vspace*{8pt}
\caption{Plot of $\vert \psi\vert^2$ varying the variables $u,\bar{v}$ and $\theta=4$. \protect\label{0-3}}
\end{figure}

\begin{figure}[ph]
\centerline{
\includegraphics[width=0.45\textwidth]{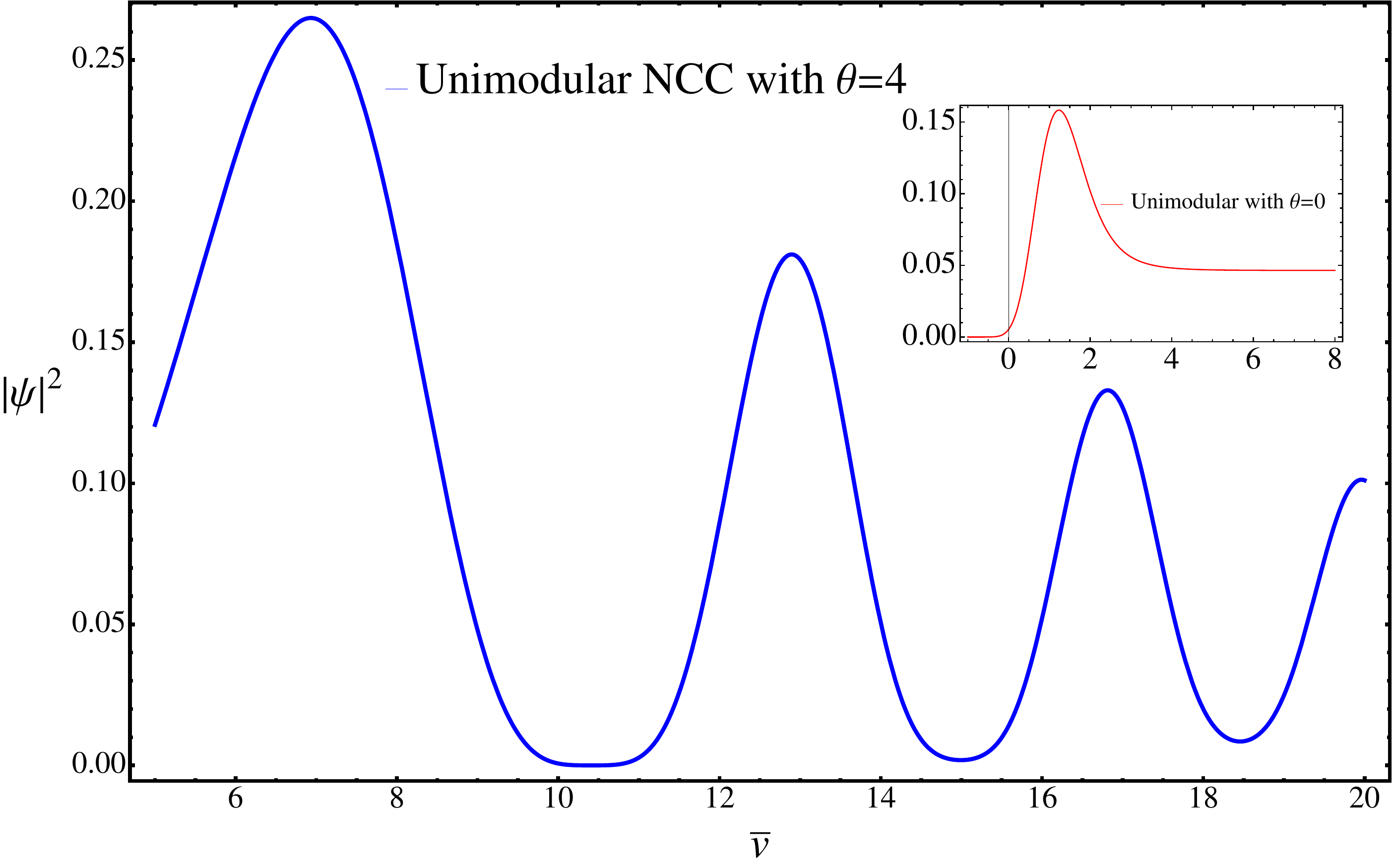}\label{pd4}}
\hspace*{8pt}
\caption{Variation of $\vert \psi\vert^2$ along the main peaks with $\theta=4$. The inner plot shows $\vert \psi\vert^2$ with $\theta=0$. \protect\label{0-4}}
\end{figure}

Now, looking for the effects of noncommutativity, let us consider the following wave packet weighted by a Gaussian
\begin{equation} 
\psi=\int^{\infty}_{-\infty} e^{-a(p_1-b)}\psi_{nc}dp_1\label{den}
\end{equation}
given the dependence of $\psi_{nc}$ on the parameter $p_1$. Since the scenario where the influence of the parameter $\theta$ could play a prominent role is in the quantum behavior of the early universe \cite{22}, for the UG context  FIG. \ref{0-3} shows the changes that the universe could have if noncommutativity is considered, 
and FIG. \ref{0-4} the possibility of tunneling among states where the peaks appears. The integral \eqref{den} is solved numerically for the variables $u,\bar{v}$ in the case $p_1>0$, $\lambda<\sqrt{3}$ with the parameters $a=1.5$, $b=1.3$, $c_0=0, c_1=1$ and  $\epsilon_0=\lambda=1$.

\section{Semiclassical Approximation}\label{Sec:UGSEM}

We shall make use of the semiclassical approximation to extract the dynamics of the UGWDW equation, taking in mind that such approximation hides the problem of time and hence the dynamical evolution of the minisuperspace can be obtained. The semiclassical limit is achieved considering in \eqref{wdwp} the WKB procedure with  $\psi(x,\phi)\approx e^{i(S_1(x)+S_2(\phi))}$ and the conditions
\begin{equation}
(\partial_{q^\mu}S_\mu)^2 \gg \partial^2_{q^\mu}S_\mu, 
\end{equation}
resulting the aforementioned EHJ equation
\begin{eqnarray}
\frac{1}{24}\Big(\frac{dS_1}{dx}\Big)^2-\frac{1}{2}\Big(\frac{dS_2}{d\phi}\Big)^2+\epsilon_0V_0e^{6x}e^{-\lambda\phi}+\nonumber\\
2\epsilon_0\Lambda e^{6x}+
6\epsilon_0ke^{4x}=0.\label{ejh}
\end{eqnarray}
Using the definition for the momenta
\begin{align}
\partial_{x}{S_1}&=\partial_{\dot{x}}{\mathcal{L}}=12\epsilon_0e^{6x}\dot{x}\nonumber\\
\partial_{\phi}{S_2}&=\partial_{\dot{\phi}}{\mathcal{L}}=-\epsilon_0e^{6x}\dot{\phi},\label{momenta}
\end{align}
equation \eqref{ejh} transforms in
\begin{eqnarray}
6\epsilon_0(\dot{x})^2-\frac{\epsilon_0}{2}(\dot{\phi})^2+V_0G_1(x,\phi)+
2\Lambda e^{-6x}+\nonumber\\
6ke^{-8x}=0,\label{ejh1}
\end{eqnarray}
where\footnote{The next allow us to avoid cross terms.} 
\begin{align}
G_1(x,\phi)
&=e^{-6x}e^{-\lambda\phi}\nonumber\\
&=\sum_{n\in \mathbb{N}\cup\{0\} }\frac{(-1)^{n}(6x+\lambda\phi)^{n}}{n!}\nonumber\\
&\approx1-6x-\lambda\phi,
\end{align}
we get the relations
\begin{align}
\frac{dt}{\sqrt{6\epsilon_0}}&=\frac{dx}{\sqrt{\vert C+6V_0x-2\Lambda e^{-6x}-6ke^{-8x}\vert}}\nonumber\\
\frac{\sqrt{2}dt}{\sqrt{\epsilon_0}}&=\frac{d\phi}{\sqrt{C+V_0(\lambda\phi-1)}}.\label{clas}
\end{align}
The first equation in \eqref{clas}, under the transformation $\alpha \leftrightarrow -x$, we find an implicit expression for $\alpha$ given by 
\begin{equation}
-t=\sqrt{12\epsilon_0}\int{\frac{d\alpha}{\sqrt{\vert C-12V_0\alpha-4\Lambda e^{6\alpha}-12ke^{8\alpha}\vert}}},\label{alpha}
\end{equation}
and the second one is straightforward to solve for the variable $\phi$
\begin{equation}
\phi=\frac{\lambda V_0}{4}\Big(\frac{\sqrt{2}t}{\sqrt{\epsilon_0}}+C_1\Big)^{2}-\frac{C}{\lambda V_0}+\frac{1}{\lambda}.\label{phi}
\end{equation}
In the same manner, \eqref{nccwdwp} with  \eqref{trans}, \eqref{relat} and 
$\psi_{nc}(u,v)\approx e^{i[\pm p_1u+f(\theta)]} e^{iS_1(v)}$, where $f(\theta)$ shows the contribution of the parameter of no commutation, gives for large $u$
\begin{eqnarray}
\Big(\frac{d S_1}{dv}\Big)^2-3(\lambda^2-3)p_1^2+\epsilon_0\Big(12\Lambda e^{3v} e^{\pm 3p_1\lambda\theta}+\nonumber\\
36ke^{2v}e^{\pm 2p_1\lambda\theta} \Big)=0.\label{nocc}
\end{eqnarray}
As a final step, in \eqref{nocc} applying \eqref{momenta}, we are in position to derive a NCC relation (implicit for $\alpha$)\footnote{Considering a first order approximation for $e^{-12x}$ and the transformation $\alpha \leftrightarrow -x$.}
\begin{eqnarray}
-t=\sqrt{12\epsilon_0}\int{\frac{d\alpha}{\sqrt{\vert \mathcal{M}\vert}}},\label{alphancc}
\end{eqnarray}
with $\mathcal{M}=C+12\kappa^2\alpha-4\Lambda e^{6\alpha}e^{\pm 3p_1\lambda\theta}-12ke^{8\alpha}e^{\pm 2p_1\lambda\theta}$ and $\kappa=\sqrt{\lambda^2-3}p_1$. It is interesting to note in the last term that when $\lambda=2$, $p_1=\sqrt{V_0}$ and $\theta \to 0$ we get \eqref{alpha}.
In this vein, when $\kappa\to 0$ and $k=0$, in the UG NCC semiclassical relation\footnote{If $\theta$ tends to zero we get the respective commutative relation.} for $\alpha$ we deduce, with $\Lambda_{nc}=4\Lambda e^{\pm3\lambda\theta}$
 \begin{equation}
a^{3}_{nc}(t)\approx(\Lambda_{nc})^{-\frac{1}{2}}\Big[\rm sech\Big(\frac{\sqrt{3}t}{2}\Big)\Big],
\end{equation}
being a similar expression in a flat space for the scale factor derived for the respective NCC solution in GR \cite{23}. Also, if CC does not  contribute and $V_0$ is small enough, there are no modifications.

\section{Conclusions and Outlooks}\label{CO}

Here we investigated Noncommutative Quantum Cosmology for the FLRW model in the Unimodular context, although QC presents a limited model to describe the quantum theory of the universe. Exists in literature a proposal for a noncommutative theory of gravity \cite{24,25},  our work has provided a picture, as seen in previous references \cite{22,26} for GR, that shows the drastically change at the time of considering noncommutativity in the early Universe using Unimodular Gravity. We find that the construction and computation of the NCC-WDW equation is a complicated task, due at each $\theta$ order, higher derivates will appear. We compare the states with those obtained in the noncommutative minisuperspace and the parameter of noncommutativity proposed, together with the case $p_1>0, \lambda<\sqrt{3}$, are selected for a better approach, observing that making $c_0=0$ gives a similar expression already reported \cite{22}. In the same manner that in GR, the deformation of the minisuperspace variables creates new possible states of the Universe evolving not only from the state in the commutative case, but from any other represented in FIG. \ref{0-3} where the peaks compete to be the most probable state of the Universe. On the other hand, the semiclassical case in the UG framework shows a commutative and NCC expression for the variable $\alpha$ is quite similar to the respective one in GR but we must note that the evolution in time for the variable $\phi$ is different, since the UG expression differs with the solution found in GR, without affecting the desire to study some cosmological parameters. In general, the wave functions $\psi_{nc}$ and the equation \eqref{alphancc}, under some considerations in the parameters, show equivalences in the QC formalism between UG and GR for the model.

\begin{acknowledgments}
J.A. A.-M. acknowledges CONACyT for the support by a postdoctoral fellowship at CINVESTAV, M\'exico. M.A.G.-A. acknowledges support from c\'atedra Marcos Moshinsky and Universidad Iberoamericana for support with the SNI grant.
E.A. M.-B.  is grateful to M. Sabido for discussions and very useful comments and Adler, also the authors recognize the useful comments of the reviewers.

\end{acknowledgments}


\begin{thebibliography}{0}    

\bibitem{1} A.G. Riess, V. Filippennko, P. Chalis, A. Clocchiatti, A. Diercks, et al, {\it The Astronomical Journal. } {\bf 116}, 1009,  (1998).

\bibitem{2} N. Aghanim, et al (Planck), arXiv:1087.06209 [astro-ph.CO].

\bibitem{3} M. Carroll, {\it Living Rev. Rel.} {\bf 4}, 1,  (2001), arXiv:astro-ph/004075 [astro-ph.CO].

\bibitem{4} S. Weinberg, {\it Reviews of Modern Physics} {\bf 61}, (1989).

\bibitem{5} Y. B.  Zeldovich, {\it Soviet Physics Uspekhi} {\bf 11}, (1968).

\bibitem{6} A.  Hern\'andez-Almada, J. Magana, M.A. Garc\'ia-Aspeitia and V. Motta, {\it Eur. Phys. J.C.} {\bf 79} 12, (2019), arXiv:1805.07895 [astro-ph.CO].

\bibitem{7} A.  Hern\'andez-Almada, G. Leon, J. Magana, M.A. Garc\'ia-Aspeitia and V. Motta, {\it Mon. Not. Roy. Astron. Soc.} {\bf497} 1590, (2020), arXiv:2002.12881 [astro-ph.CO].

\bibitem{8} R. Maartens, {\it Living Rev. Rel.} {\bf 7}, 7,  (2004), arXiv:gr-qc/03120 [gr-qc].

\bibitem{9}M.A. Garc\'ia-Aspeitia,  A.  Hern\'andez-Almada, J. Magana, M.H. Amante, V. Motta and C. Mart\'ines-Robles, {\it Phys. Rev. D} {\bf97} 101301, (2018), arXiv:1804.05085 [gr-qc].

\bibitem{10}M.A. Garc\'ia-Aspeitia,  J. Magana, A.  Hern\'andez-Almada and V. Motta, {\it Phys. Rev. D} {\bf27} 1850006, (2017), arXiv:1619.08220 [astro-ph.CO].

\bibitem{11}C. Escamila-Rivera, A.  Hern\'andez-Almada, M.A. Garc\'ia-Aspeitia and V. Motta, (2020), 10.1142/S0218271821500772, arXiv:2005.13957 [gr-qc].

\bibitem{12}M.A. Garc\'ia-Aspeitia, C. Mart\'ines-Robles, A.  Hern\'andez-Almada, J. Magana and V. Motta, {\it Phys. Rev. D} {\bf99} 123525, (2019), arXiv:1903.06344 [gr-qc].

\bibitem{13}M.A. Garc\'ia-Aspeitia, A.  Hern\'andez-Almada, J. Magana and V. Motta, {\it Phys. Dark Univ.} {\bf32} 100840, (2021), arXiv:1912.07500 [astro-ph.CO].

\bibitem{14}V. Motta, M.A. Garc\'ia-Aspeitia, A.  Hern\'andez-Almada, J. Magana and T. Verdugo, {\it Universe} {\bf7} 163, (2021), arXiv:2104.04642 [astro-ph.CO].

\bibitem{15}G. F. Ellis, H. van Elst, J. Murugan and J.P. Uzan, {\bf28} 225007, (2011).

\bibitem{16}A. Eichhorn, {\it Class. Quant. Grav.} {\bf30} 115016, (2013), arXiv:1301.0879 [gr-qc].

\bibitem{7} A. H. Chamseddine, Phys. Lett. B 504, 33 (2001).
\bibitem{8} O. Obregon, M. Sabido, E. Mena, Mod. Phys. Lett. A24, 1907-1914 (2009).

\bibitem{17}L. Smolin, {\it Phys. Rev. D.} {\bf80} 084003, (2009), arXiv:0904.4841 [hep-th].

\bibitem{18}W. Guzm\'an, M. Sabido, J. Socorro and L. Urena-L\'opez, {\it Int. J. Mod.Phys. D} {\bf 16}, 641, (2007), arXiv:gr-qc/0506041.

\bibitem{19}arXiv:hep-th/0410213.

\bibitem{gamboa}J. Gamboa, M. Loewe and J. C. Rojas, Phys. Rev D {\bf 64}, 067901 (2001); M. Chaichian, M. M. Sheikh-Jabbari and A. Tureanu, Phys. Rev. Lett. {\bf 86} 2716 (2001).

\bibitem{20}E. Alvarez, S. Gonzalez-Mart\'in, M. Herrero-Valea and C.-Mart\'in, {\it JHEP} {\bf 78}, 220151, (2015), 10.1007/JHEP08(2015)078, arXiv:1505.01995.

\bibitem{21}J.A. Astorga-Moreno, J. Chagoya, J. Flores-Urbina and M. A. Garc\'ia-Aspeitia, {\it JCAP} {\bf 09}, (2019), 10.1088/JCAP09(2019)005, arXiv:1905.11253.

\bibitem{22}H. Garc\'ia-Compe\'an, O. Obreg\'on and C. Ram\'irez, {\it Phys.  Rev. Lett.} {\bf 88}, (2002).

\bibitem{23}E. A. Mena-Barboza, O. Obreg\'on and M. Sabido, {\it Int. J. Mod. Phys. D} {\bf 18}, (2015).

\bibitem{24}N. Seiberg and E. Witten, {\it JHEP} {\bf 09}, 032, (1999).

\bibitem{25}N. Nekrasov and A. Schwarz, {\it Commun. Math. Phys.} {\bf 198}, 689, (1999).

\bibitem{26}A. Crespo-Hernandez, E. A. Mena-Barboza, {\it Int. J. Mod. Phys. D} {\bf 29}, (2020).

\end{thebibliography}
\end{document}